\documentclass[superscriptaddress,pra,reprint,twocolumn,amsmath,amssymb,floatfix]{revtex4-2}
\usepackage{multirow}
\usepackage{graphicx}
\usepackage{booktabs}
\usepackage{natbib}
\usepackage[colorlinks=true, linkcolor=blue, citecolor=blue, urlcolor=blue]{hyperref} % Enable and customize hyperlinks
\usepackage{tabularx}
\usepackage{bm}
\usepackage{comment}
\usepackage{harpoon}
\usepackage{makecell}
\usepackage{subfigure}
\usepackage{verbatim}
\usepackage{orcidlink}  
\allowdisplaybreaks
\begin{document}

\begin{abstract}
Metastability-exchange collisions (MECs) lie at the heart of metastability-exchange optical pumping (MEOP) in $^3$He, enabling the transfer of polarization from the metastable state to the ground state, as well as the optical detection of nuclear magnetic resonance. 
Leveraging MECs, optically pumped $^3$He nuclear magnetometers have been developed since the earliest demonstrations of MEOP. 
However, it also induces an additional frequency shift and relaxation of the nuclear spin precession, thereby limiting the sensitivity of the magnetometer.
In this work, we identify a new source of frequency shift and relaxation in the $^3$He nuclear spin, arising from the light shift.
This effect arises from an MEC-mediated interaction between light and the nucleon spin.
We develop a theoretical model to describe this light-induced effect and highlight its significance in low magnetic fields.
This effect is experimentally demonstrated, and its dependence on various parameters—including magnetic field strength, light intensity, and wavelength—is investigated.
Our result provides a better understanding of the frequency shift and relaxation of $^3$He spin precession under MEOP conditions.
Moreover, our experiment reveals an MECs-mediated coupling between the $^3$He nuclear spin and light, which may indicate the feasibility of MECs-assisted optical manipulation of $^3$He nuclear spins at the quantum level, as proposed in several theoretical schemes.
\end{abstract}

\title{Light-induced Frequency Shift and Relaxation of Ground-State $^3$He via Metastability-Exchange Collisions}

\author{L. Y. Wu \orcidlink{0009-0003-7568-008X}}
\email[]{22110200026@m.fudan.edu.cn}
\affiliation{Key Laboratory of Nuclear Physics and Ion-beam Application (MOE), Institute of Modern Physics, Fudan University, Shanghai 200433, China}

% \author{C. Fu \orcidlink{0000-0001-5895-7823}}
% \email[Contact author: ]{cbfu@fudan.edu.cn}
% \affiliation{Key Laboratory of Nuclear Physics and Ion-beam Application (MOE), Institute of Modern Physics, Fudan University, Shanghai 200433, China}

\author{H. Yan \orcidlink{0000-0002-8286-4691}}
\email[Contact author: ]{yanhaiyang@nbu.edu.cn}
\affiliation{Institute of Fundamental Physics and Quantum Technology, and School of Physical Science and Technology, Ningbo University, Ningbo, Zhejiang 315211, China}
\maketitle

\section{Introduction}
\label{sec:1}
The study of metastability-exchange optical pumping (MEOP) of $^3$He dates back to the 1960s \cite{Walters1962}. 
Due to its high efficiency and ability to achieve high $^3$He polarization, the MEOP technique has become one of the primary methods for polarizing $^3$He, with applications in nuclear physics \cite{Anthony1996prd}, magnetic resonance imaging \cite{Karpuk2013PPN}, nuclear spin magnetometry or gyroscopes \cite{Gemmel2010EPJD, walker2016}, and fundamental physics testing \cite{gentile2017RMP, safronova2018, cong2024arxiv, Zhang2025prl}. Since the ground state of $^3$He is inaccessible to laser excitation due to an energy gap of approximately 20 eV to the first excited state, the MEOP technique relies on polarizing the metastable state with optical pumping and transferring the polarization to the ground state via metastability-exchange collisions (MECs). 

The MECs play a crucial role in polarizing the ground state of $^3$He by transferring polarization from the metastable state to the ground state \cite{nacher1985JP}. 
The high efficiency of the MEOP is attributed to the fact that all involved angular momenta are spins.
Meanwhile, the MECs process is extremely fast ($\sim 10^{-7}$s), causing minimal depolarization during collisions and enabling efficient transfer of angular momentum from photons to the nuclear spin \cite{nacher1985JP}. 
In addition, the rapid MECs process establishes a strong coupling between the ground and metastable states, facilitating coherence exchange between them \cite{partridge1966pps}.
In the early development of MEOP, this property was exploited to detect magnetic resonance in the ground state by probing the metastable state with a laser \cite{dupont1973b, dupont1973JP,colegrove1963}. An optically detected nuclear magnetometer utilizing MEOP-polarized $^3$He was developed by Schearer \textit{et al.} for measuring the Earth's magnetic field with a sensitivity of several microgauss \cite{Schearer1963RSI}. Recently, MECs have been leveraged to observe the dressing effect of the nuclear spin \cite{zhan2019a, zhan2019OE}. 

Conversely, the ground state is also affected by the metastable state through MECs. 
Both the Larmor frequency and the relaxation time of the nuclear spin are modified by MECs, potentially limiting the accuracy of magnetometers \cite{Dehmelt1964RSI, Schearer1964RSI}. 
In the realm of quantum information, MECs have been proposed as a mechanism to mediate effective interactions between $^3$He nuclear spins and light, enabling the generation of spin-squeezed states \cite{Dantan2005PRL, Serafin2021PRL, Fadel2024NJP, Katz2024arXiv}. 
An effective Faraday interaction, $H\propto I_zS_z$, where $I_z$ is the spin of the nucleus and $S_z$ is the Stokes spin operator of the light, is yielded in the nondemolition measurement spin squeezing scheme \cite{Serafin2021PRL, Fadel2024NJP}.
Whereas, evidence of nuclear spin-light coupling has not been observed in experiments yet.
In this work, we identify a new source of frequency shift and relaxation over the ground state $^3$He in MEOP induced by the pump light.
We develop a theoretical model to describe this light-induced effect and find good agreement with the experiment.
Given that the Faraday interaction between light and atoms also leads to a shift in the spin precession frequency, our experimental results provide practical support for MECs-mediated, optically driven nuclear spin control.
The paper is organized as follows.
In Sec. \ref{sec:2}, a model that describes the frequency shift and relaxation of the $^3$He nuclear spin induced by the light is introduced. In Sec. \ref{sec:3}, we demonstrate the effect experimentally under various conditions and compare the results with theoretical predictions to validate the model.
Section \ref{sec:4} presents the conclusion of this paper.

\section{Theoretical Model}
\label{sec:2}
\subsubsection{Light Shift on the metastable state}

We first consider the effect of the pump light on the metastable state of $^3$He. The pump light propagating along the $z$-axis is represented by a monochromatic wave
\begin{equation}
    \boldsymbol{E}(t)=\frac{1}{2}E_0(\boldsymbol{\epsilon}e^{-i\omega t}+\boldsymbol{\epsilon}^*e^{i\omega t}),    
\end{equation}
where $\omega$ is angular frequency, $E_0$ is the amplitude of the electric field, and $\boldsymbol{\epsilon}$ is the complex polarization vector. 
Since the typical light intensity ($100$ mW/cm$^2$) used in our experiment to address the $2^3S_1$-$2^3P_0$ transition is much lower than the saturation intensity ($\sim$1 W/cm$^2$), the population in the excited state can be adiabatically 
eliminated. 
Consequently, the influence of the light can be captured by an effective Hamiltonian involving only the ground-state variables, taking the form \cite{happer1967PR,lee2019a}:
\begin{equation}
\label{eq1}
    \delta\mathcal{H}=-\frac{|E_0|^2}{4}\boldsymbol{\epsilon}^*\cdot\overleftrightarrow{\alpha}\cdot\boldsymbol{\epsilon},
\end{equation}
where $\overleftrightarrow{\alpha}$ is the complex polarizability tensor operator. 
In general, the effective Hamiltonian $\delta\mathcal{H}$ is not Hermitian and can be decomposed into a real part $\delta \hat{\mathcal{E}}=(\delta\mathcal{H}+\delta\mathcal{H}^\dagger)/2$ representing the light-shift operator, and an imaginary part $\delta \hat{\Gamma}=i(\delta\mathcal{H}-\delta\mathcal{H}^\dagger)$ representing the light-absorption operator. 
Therefore, the evolution of the density matrix of the metastable state under the effective Hamiltonian reads
\begin{equation}
\label{dens}
    i\frac{d\rho_m}{dt}=[H_0+\delta\hat{\mathcal{E}},\rho_m]-\frac{i}{2}(\delta \hat{\Gamma}\rho_m+\rho_m\delta \hat{\Gamma}),
\end{equation}
where $H_0$ is the Hamiltonian in the absence of light, including the hyperfine interaction and the Zeeman interaction.
This indicates that the light-shift operator shifts the atomic energy levels, while the light-absorption operator accounts for dissipative effects caused by the loss of atoms due to light absorption.
In a Zeeman multiplet of the metastable state, the influence of the light results in a shift of the resonance frequency and broadening of the linewidth. 

The effective Hamiltonian $\delta\mathcal{H}$ is constructed from the scalar product of two rank-2 tensors and can therefore be decomposed into a sum of scalar products involving irreducible tensor components. 
Specifically, in terms of the hyperfine levels in the metastable state of $^3$He, the vector part of the effective Hamiltonian $\delta\mathcal{H}$ can be written as
%begin{widetext} 
\begin{equation}
\label{eq2}
\begin{aligned}
    \delta\mathcal{H}^{(1)}&=\frac{Ie^2f_{ge}}{4 c\varepsilon_0m_e}\boldsymbol{s}\cdot\Big(\frac{2Z(\omega;\omega_8)}{\omega_8}\boldsymbol{F}_{1/2}+\frac{Z(\omega;\omega_9)}{\omega_9}\boldsymbol{F}_{3/2}\Big),
\end{aligned}
\end{equation}
%\end{widetext}
where $I=\frac{1}{2}c\varepsilon_0|E_0|^2$ is the intensity of the laser, $c$ is the speed of light, $\varepsilon_0$ is the vacuum permittivity, $\omega_8$ ($\omega_9$) is the $C_8$ ($C_9$) transition frequency of the hyperfine levels, $f_{ge}=0.0599$ is the oscillator strength of $2^3S_1-2^3P_0$ transition \cite{Wiese2009}, and $\boldsymbol{s}=(\boldsymbol{\epsilon}^*\times\boldsymbol{\epsilon})/i$ is the mean photon spin determined by the polarization of the pump laser. 
The plasma dispersion function $Z(\omega;\omega_0)$ is defined as
\begin{equation}
\begin{aligned}
     Z(\omega;\omega_0)&=\frac{1}{D\sqrt{\pi}}\int^{\infty}_{-\infty}\frac{e^{-u^2}}{u-\frac{\omega_0-\omega+i\gamma_e}{D}}du\\
     &=D^{-1}i\sqrt{\pi}w(\frac{\omega_0-\omega+i\gamma_e}{D}).
\end{aligned}
\end{equation}
Here $\gamma_e$ is the radiation decay of the excited state $2^3P_0$, $w(x)=e^{-x^2}[1+\mathrm{erf}(ix)]$ is the complex error function, and $D={\omega_0}\sqrt{\frac{2k_BT}{m_{\rm He}c^2}}$ is the Doppler linewidth where $k_B$ is the Boltzmann constant and $m_{\rm He}$ is the mass of $^3$He. 
The function $Z(\omega;\omega_0)$ is, in fact, a convolution of a Lorentzian profile with the Maxwell velocity distribution, representing the atomic response to the laser. 
Note that the coherence between the two hyperfine levels in the effective Hamiltonian is neglected due to the existence of the MECs process \cite{courtade2002EPJD}.
In the Hamiltonian (\ref{eq2}), we retain only the vector component relevant to our discussion.
A detailed derivation of the effective Hamiltonian is provided in Appendix \ref{app1}. 
The effect of $\delta\mathcal{H}^{(1)}$ clearly resembles an interaction between a spin and a magnetic field, which will cause a shift to the magnetic resonance frequency of the electron. Substituting $\delta\mathcal{H}^{(1)}$ into Eq. (\ref{dens}), the evolution of the transverse angular momentum in the metastable state is given by:
\begin{equation}
\label{eq6}
    \begin{aligned}
        \frac{\rm d}{\rm dt}F_{3/2,+} &=i\frac{2}{3}\gamma_m( B_0+\delta B_{3/2}) F_{3/2,+},\\
        \frac{\rm d}{\rm dt}F_{1/2,+} &=i\frac{4}{3}\gamma_m( B_0+\delta B_{1/2})F_{1/2,+}.\\
    \end{aligned}
\end{equation}
Here, $B_0$ is the magnetic field along $z$-axis, $\gamma_m$ is the gyromagnetic ratio of the electron, $F_{3/2,+}$ ($F_{1/2,+}$) is the transverse angular momentum of the $F=3/2$ ($F=1/2$) state, and the effective fields $\delta B_{3/2}$ and $\delta B_{1/2}$ are 
\begin{equation}
    \begin{aligned}
        \delta B_{1/2} &=\frac{3f_{eg}Ie^2}{8\hbar\gamma_mc\varepsilon_0m_e}\frac{Z(\omega;\omega_8)}{\omega_8}s_z, \\
        \delta B_{3/2}&=\frac{3f_{eg}Ie^2}{8\hbar\gamma_mc\varepsilon_0m_e}\frac{Z(\omega;\omega_9)}{\omega_9}s_z,\\
    \end{aligned}
\end{equation}
in which the real part corresponds to the frequency shift, and the imaginary part accounts for the relaxation.

\subsubsection{The metastability-exchange collisions}

Under discharge conditions, the metastable state of $^3$He is strongly coupled to the ground state via the MECs. 
It has been found that the coherence or transverse polarization of the nuclear spin can be transferred to atoms in the metastable state through the MECs process, provided that the Larmor frequency in the metastable state is lower than the MECs rate \cite{partridge1966pps}. Additionally, the frequency and relaxation rates in the ground state are also modified by the MECs \cite{Dehmelt1964RSI,dupont1973JP,dupont1973b}.
The evolution of the ground state and metastable state during the MECs could be described as follows \cite{dupont1973JP}:
\begin{equation}
    \begin{aligned}
        \frac{\rm d\rho_g}{\rm dt}&=\frac{1}{T}(-\rho_g+\mathrm{Tr_e}[\rho_m]),\\
        \frac{\rm d\rho_m}{\rm dt}&=\frac{1}{\tau}(-\rho_m+\sum_{F=1/2,3/2}P_F(\mathrm{Tr_n}[\rho_m]\otimes\rho_g)P_F),
    \end{aligned}
\end{equation}
where $\rho_g$ is the density matrix of the ground state, $T$ (typically 1 s) and $\tau$ (typically $10^{-7}$ s) are the average times of the MECs process for the ground and metastable states, respectively, $P_F$ is the projector operator onto the hyperfine level, and $\mathrm{Tr_n}$ and $\mathrm{Tr_e}$ denote the trace operators over the nuclear and electronic variables, respectively. 
The two collision times $T$ and $\tau$ are related by the relationship $\tau/T=n/N$ where $n$ and $N$ are the number densities of metastable and ground state atoms, respectively. 
% This relationship ensures conservation of total angular momentum during collisions.
The evolution of the transverse angular momentum in the ground and metastable states under the MECs could be written as~\cite{dupont1973JP}
\begin{equation}
\label{eq9}
\begin{aligned}
        \frac{\rm d}{\rm dt}I_{+}&= -\frac{1}{T}I_++\frac{1}{3T}(F_{3/2,+}-F_{1/2,+}),\\
        \frac{\rm d}{\rm dt}F_{3/2,+}&=-\frac{4}{9\tau}F_{3/2,+}+\frac{10}{9\tau}F_{1/2,+}+\frac{10}{9\tau}I_+,\\
        \frac{\rm d}{\rm dt}F_{1/2,+}&=-\frac{7}{9\tau}F_{1/2,+}+\frac{1}{9\tau}F_{3/2,+}-\frac{1}{9\tau}I_+,
\end{aligned}
\end{equation}
where $I_+$ is the transverse angular momentum of the ground state.
Notably, although the angular momentum in different levels varies with time, the total angular momentum, given by $NI_++n(F_{3/2,+}+F_{1/2,+})$, remains conserved throughout the evolution. 
This is the result of the spin-independent nature of the MECs process. 
In the steady state, it also shows that nuclear polarization in the ground state $I_{+}$ is equal to that in the metastable state $I_{m,+}=1/3(F_{3/2,+}-F_{1/2,+})$.
Furthermore, due to the condition $\tau \ll T$, the MECs process enforces an adiabatic following of the angular momentum in the metastable state by that in the ground state.
A simple illustration is provided here, with a more detailed explanation given in the following section.
On the timescale of approximately $\tau$, the nuclear spin remains effectively frozen owing to the condition $T \gg \tau$. 
In contrast, the angular momentum in the metastable state evolves on a timescale of $1/\tau$, allowing it to align instantaneously with the direction of the nuclear spin. 
Consequently, when the nuclear spin precesses in a magnetic field at frequency $\omega$, the angular momentum in the metastable state will precess at the same frequency, as long as the condition $\omega \tau \ll 1$ is satisfied.

\subsubsection{Frequency shift and relaxation of the ground state}

We now consider the magnetic resonance of ground-state $^3$He in the presence of the MECs. 
Combining Eqs.~(\ref{eq6}) and (\ref{eq9}), we obtain the evolution equations for the transverse angular momenta of the ground and metastable states as follows:
% \begin{widetext}
{\small
\begin{equation}
\label{eqtrang}
    \begin{aligned}
        \frac{\rm d}{\rm dt}I_+&=[i\gamma_gB_0-(\frac{1}{T}+\frac{1}{T_r})]I_++\frac{1}{3T}(F_{3/2,+}-F_{1/2,+}),\\
        \frac{\rm d}{\rm dt}F_{3/2,+}&=\frac{10}{9\tau}(I_++F_{1/2,+})+[i\frac{2}{3}\gamma_m(B_0+\delta B_{3/2})-\frac{4}{9\tau}]F_{3/2,+},\\
        \frac{\rm d}{\rm dt}F_{1/2,+}&=\frac{1}{9\tau}(F_{3/2,+}-I_+)+[i\frac{4}{3}\gamma_m(B_0+\delta B_{1/2})-\frac{7}{9\tau}]F_{1/2,+},
    \end{aligned}
\end{equation}}
% \end{widetext}
% {\small
% \begin{widetext}
% \begin{equation}
% \label{eqtrang}
% \frac{\rm d}{\rm dt}\left [\begin{array}{c}
%      I_{+} \\
%      F_{3/2,+} \\
%      F_{1/2,+}
% \end{array}\right ]=
% \left [\begin{array}{ccc}
%         i\gamma_gB_0 -(\frac{1}{T}+\frac{1}{T_r}) & \frac{1}{3T} & -\frac{1}{3T}\\
%         \frac{10}{9\tau} & i\frac{2}{3}\gamma_m(B_0+\delta B_{3/2}) -\frac{4}{9} & \frac{10}{9\tau}\\
%         -\frac{1}{9\tau} &  \frac{1}{9\tau} & i\frac{4}{3}\gamma_m(B_0+\delta B_{1/2})-\frac{7}{9\tau}
%     \end{array}\right ]
% \left [\begin{array}{c}
%      I_{+} \\
%      F_{3/2,+} \\
%      F_{1/2,+}
% \end{array}\right ],
% \end{equation}
% \end{widetext}}
where $\gamma_g$ is the gyromagnetic ratio of the $^3$He nucleus and $T_r$ denotes the relaxation time of the spin.
Equation (\ref{eqtrang}) describes the evolution of one slow mode with frequency $\gamma_gB_0$ and two fast modes with frequencies $2/3\gamma_m(B_0+\delta B_{3/2})$ and $4/3\gamma_m(B_0+\delta B_{1/2})$ coupled by the MECs. 
% The solution of Eq. (\ref{eqtrang}) can be obtained using the perturbation method \cite{\rm dupont1973JP}. 
The evolution of the ground state can be decoupled by adiabatically eliminating the metastable manifold.
Since the Larmor frequency of the ground state at low fields is much slower than the MECs rate $1/\tau$ and the Larmor frequencies of metastable states, the variations of these two fast modes could be treated as quasi-static at the timescale of nuclear spin evolution.
Therefore, by solving the algebraic equations $dF_{3/2,+}/dt = 0$ and $dF_{1/2,+}/dt = 0$, and substituting the solution into the motion equation of $I_+$, we can obtain the equation only concerning the ground-state nuclear spin
\begin{equation}
\label{nuI}
    \frac{\rm d}{\rm dt}I_{+}=i(\gamma_gB_0-\frac{\mathrm{Im}[\beta]}{T})I_{+}-\Big(\frac{1}{T_r}+\frac{1+\mathrm{Re}[\beta]}{T}\Big)I_{+},
\end{equation}
where
{\small
\begin{widetext}
\begin{equation}
\label{beta}
\beta= \frac{-i \gamma_m(20 \delta B_{1/2}+\delta B_{3/2}) \tau -21 i\omega _m\tau +9}{12 \gamma_m\delta B_{1/2} \tau  (3\omega _m\tau+3\gamma _m\delta B_{3/2}\tau+2 i)+3\gamma_m\delta B_{3/2}\tau(7i+12\omega_m\tau)-9 (1-i\omega_m\tau) (1-4 i\omega_m\tau )}.
\end{equation}
\end{widetext}}
\noindent This indicates that both the frequency and relaxation time of the nuclear spin are modified by the metastable state through the MECs.
Setting the light-induced field to zero, Eq. (\ref{beta}) restores to the results obtained in Ref. \cite{dupont1973b}. 
The MECs-induced frequency shift and relaxation have been observed experimentally in Ref. \cite{dupont1973JP}.

\begin{figure}[htbp]
    \centering
    \includegraphics[width=0.95\linewidth]{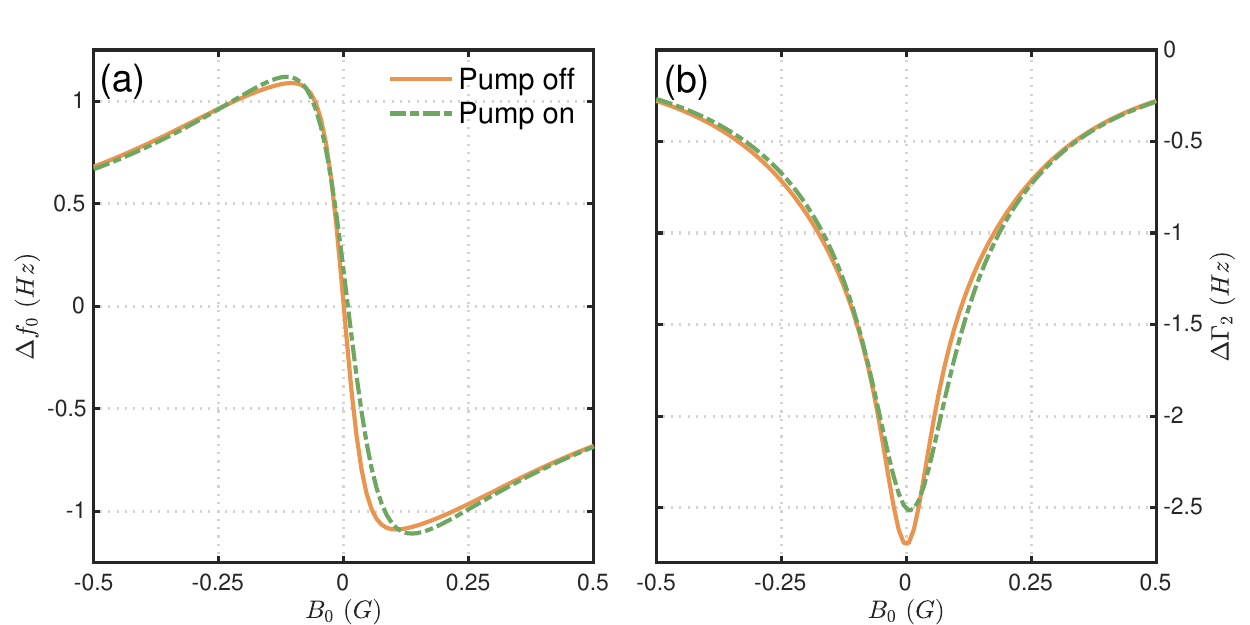}
    \caption{The MECs induced frequency shift and relaxation of the nuclear spin precession at different strengths of magnetic field calculated from Eq. (\ref{beta}). The solid and dashed lines represent the results with and without the light-shift field, respectively.}
    % The parameters used in the calculation are $\tau=2.2\times 10^{-7}\ s$, $\gamma_m=2\pi\times 2.8\times 10^6\ \mathrm{MHz/G}$ and $T=0.37\ s$.
    \label{fig:1}
\end{figure}

Here, we find that the light-induced frequency shift and relaxation in the metastable state can also be transferred to the ground state through the MECs, resulting in additional frequency shifts and relaxation.
Under typical MEOP conditions, where the pump light intensity is approximately 100 mW/cm$^2$, the light shift will induce an effective magnetic field on the order of microtesla.
As such, we expect that the influence of the pump light becomes significant in the low-field regime.
Figure \ref{fig:1} illustrates the frequency shift and relaxation rate of nuclear spin precession induced by the MECs at different magnetic field strengths.
Here, the value of the fictitious field is set to a relatively large value to clearly demonstrate the effect of the pump light on the plot.
In the high-field region, the light shift is negligible, the MECs-induced frequency shift approaches zero, and the total relaxation rate approaches $1/T_r + 1/T$.
As the magnetic field decreases, the frequency shift initially increases before subsequently diminishing to zero, while the relaxation induced by the MECs is progressively suppressed.
In the absence of the light shift, as represented by the solid curve in Fig. \ref{fig:1}(b), MECs-induced relaxation can be nearly completely suppressed near zero field, leaving nuclear spin decoherence primarily governed by the $T_r$ term.
However, when the light shift is included, as shown by the dashed curve, the relaxation time is significantly reduced, particularly in the near-zero-field region.
The light-induced relaxation arises from two factors: (i) the light generates an effective magnetic field that prevents full suppression of the MECs-induced relaxation, and (ii) the light induces relaxation in the metastable state, which subsequently contributes to an increased nuclear spin relaxation rate via MECs.
An intuitive explanation for the frequency shift and relaxation of the nuclear spin is that, during the MECs process, atoms circulate between the metastable state and the ground state.
As such, the effective resonance frequency and relaxation rate of the ground-state $^3$He are a weighted average of the contributions from the ground and metastable states.

\begin{figure}[htbp]
    \centering
        \includegraphics[width=\linewidth]{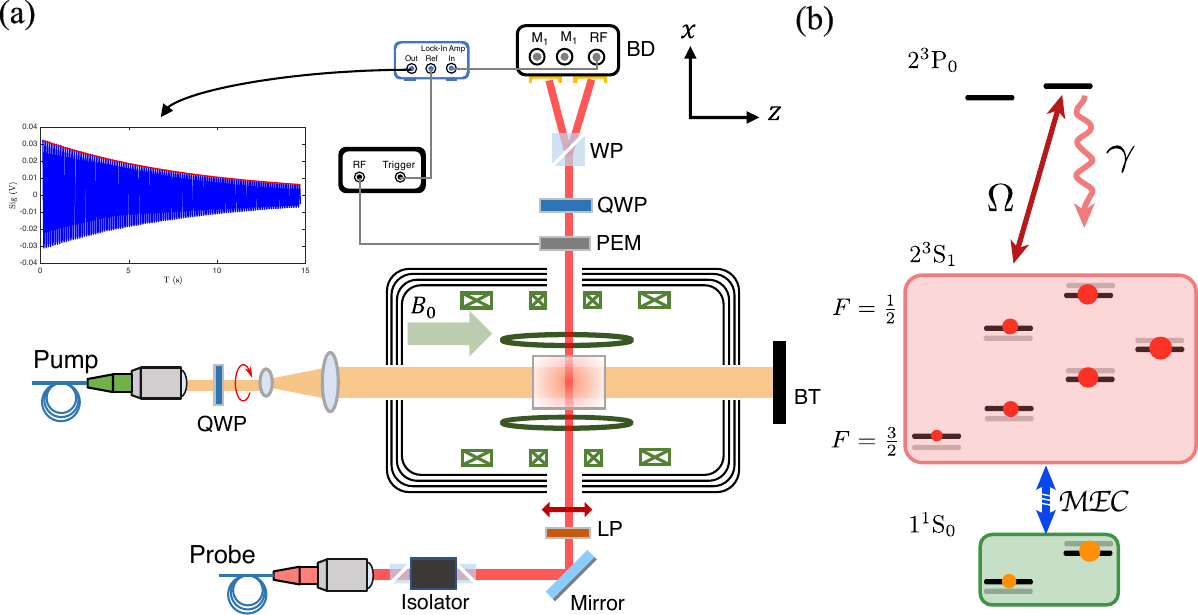}
    \caption{(a) Experimental setup of the MEOP-based $^3$He polarization system. The setup consists of three main components: (i) a guiding magnetic field generated by a Merritt coil system housed within a five-layer $\mu$-metal magnetic shield, (ii) an optical pumping system comprising a laser with approximately 2~W output power, along with polarization control and beam expansion optics, and (iii) a signal detection system employing a weak ($\sim$mW), linearly polarized probe laser propagating vertically, combined with a homodyne detection scheme. The inset plot shows a typical FID signal demodulated using a lock-in amplifier. LP: Linear Polarizer; QWP: Quarter Wave Plate; BT: Beam Trap; BD: Balanced Detector; PEM: Photoelastic Modulator; WP: Wollaston Prism. (b) Hyperfine structures of the atomic states of $^3$He involved in the optical pumping. The metastable state and ground state interact through the MECs. The red double arrow represents the optical pump light, and the wavy line indicates the decay of the excited state. The light shift alters the Zeeman sublevels in the metastable state, and through the MECs, the ground state is effected as well.}
    \label{fig:2}
\end{figure}

\section{Results and Discussion}
\label{sec:3}
We demonstrate the light-induced frequency shift and relaxation in a low-field MEOP-based $^3$He polarization system. 
The experimental setup is schematically illustrated in Fig. \ref{fig:2}. 
A small fraction of $^3$He atoms in the metastable state, excited by a radio-frequency (RF) field, is optically pumped by a circularly polarized laser. 
Through the MECs, polarization is subsequently transferred to ground-state $^3$He, thereby achieving nuclear spin polarization.
A linearly polarized weak probe light propagating along the $x$-axis is utilized to measure the nuclear spin polarization in this direction. 
The nuclear spin polarization component along the probe light direction will induce a rotation of the light’s polarization, known as the Faraday rotation effect. 
The polarization rotation angle is measured using a homodyne detection scheme. 
To enhance the signal-to-noise ratio, the signal is modulated by a photoelastic modulator and subsequently demodulated using a lock-in amplifier. 
A resonant RF pulse excites the transverse nuclear spin polarization, after which the nuclear spin precession undergoes free induction decay (FID) in the guiding magnetic field.

The Larmor frequency of the $^3$He nuclear spin is estimated by fitting the FID signal using
\begin{equation}
\label{fid}
    S_x(t)=S_0\sin(2\pi f_0t+\phi)e^{-t/T_2}+S_1e^{-t/T_1},
\end{equation}
where $2\pi f_0=\gamma_{g}B$ is the Larmor frequency and $T_2$ ($T_1$) is the transverse (longitudinal) relaxation time. The first term of Eq. (\ref{fid}) describes the coherence decay, while the second term accounts for the polarization decay. 
The second term is included because, at very low magnetic fields, the residual field inside the shield ($\sim$10 nT), which is not perfectly aligned with the $z$-axis, alters the total field direction and gives rise to nonzero longitudinal polarization along the $x$-axis.

\begin{figure}[htbp]
    \centering
    \includegraphics[width=0.9\linewidth]{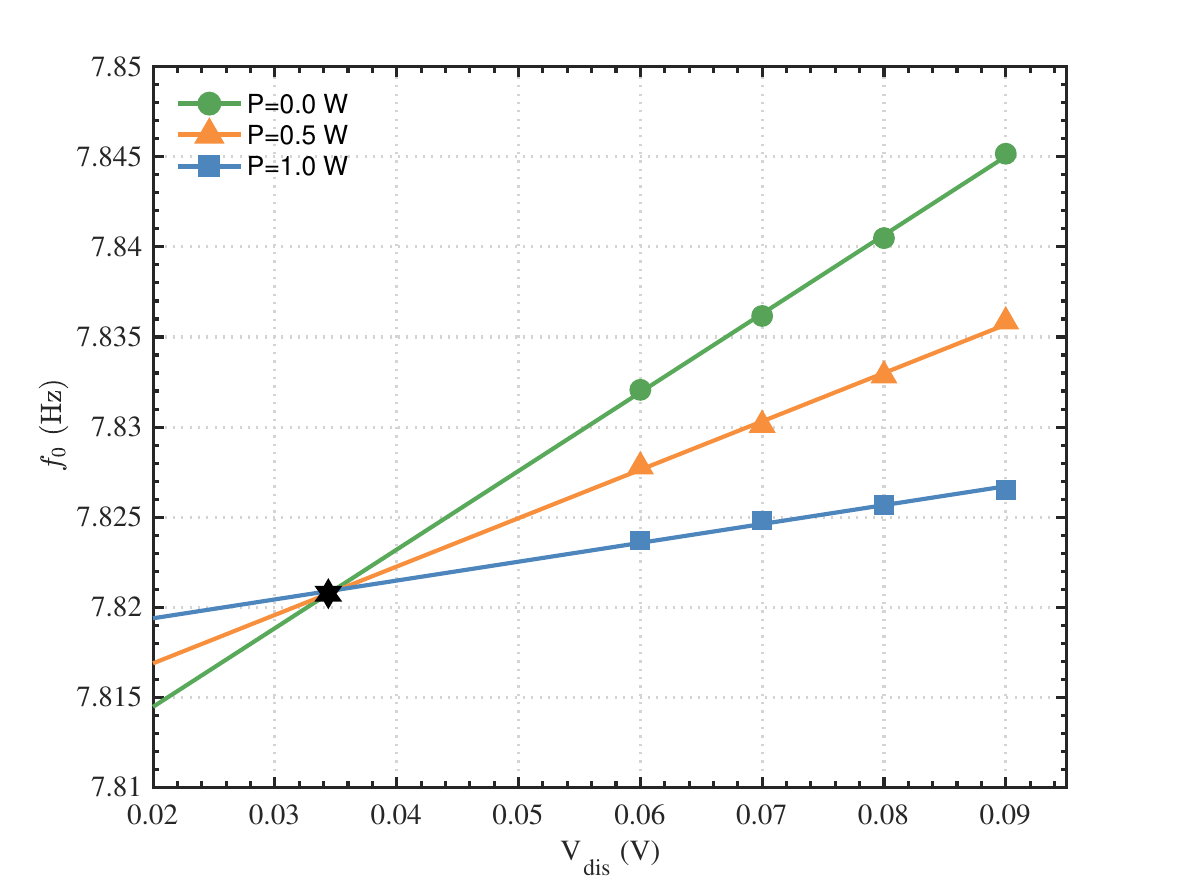}
    \caption{The Larmor precession frequency versus the strength of the radio-frequency discharge at different powers of the pump laser. The solid lines are the linear fits. The data marked in green circles, yellow triangles, and blue squares are experimental results. The black star marks the intersection of these three curves.}
    \label{fig:3}
\end{figure}

Figure \ref{fig:3} shows the frequency shift at different RF discharge strengths. 
The green circles correspond to measurements with optical pumping blocked, while the yellow triangles and blue squares correspond to measurements with optical pumping enabled.
The $V_\mathrm{dis}$ represents the amplitude of the signal generator, which is proportional to the RF power imposed in the $^3$He cell. 
As indicated by Eq. (\ref{nuI}), the frequency shift increases with the discharge strength, since the MECs rate $1/T$ is proportional to the number density $n$ of metastable atoms, which itself grows with discharge strength.

On the other hand, according to Eq. (\ref{nuI}), the slope of the curves in Fig. \ref{fig:3} depends on $\mathrm{Im}[\beta]$.
Therefore, when the pump laser is enabled, the value of $\mathrm{Im}[\beta]$ is slightly different from that of the pump-off condition. 
By extrapolating the three lines to their intersection, we obtain the ground-state Larmor frequency, unaffected by both MECs and light shift, corresponding to $\omega_g = 2\pi \times 7.821$ Hz at $V_\mathrm{dis} = 0.035$ V, where the discharge is extinguished.
% Theoretically, the three curves should have the same intercept at $V_\mathrm{dis} = 0$ V, since the MECs rate vanishes when the discharge is extinguished; however, experimental results show that they intersect at approximately $V_\mathrm{dis} = 0.035$ V.
% This discrepancy occurs because the discharge requires a nonzero ignition voltage. 
% The intersection of these three curves yields a Larmor frequency unaffected by the MECs and light shift of $\omega_g = 2\pi \times 7.821$ Hz.

\begin{figure}[htbp]
    \centering
    % \begin{subfigure}{}
    \includegraphics[width=0.9\linewidth]{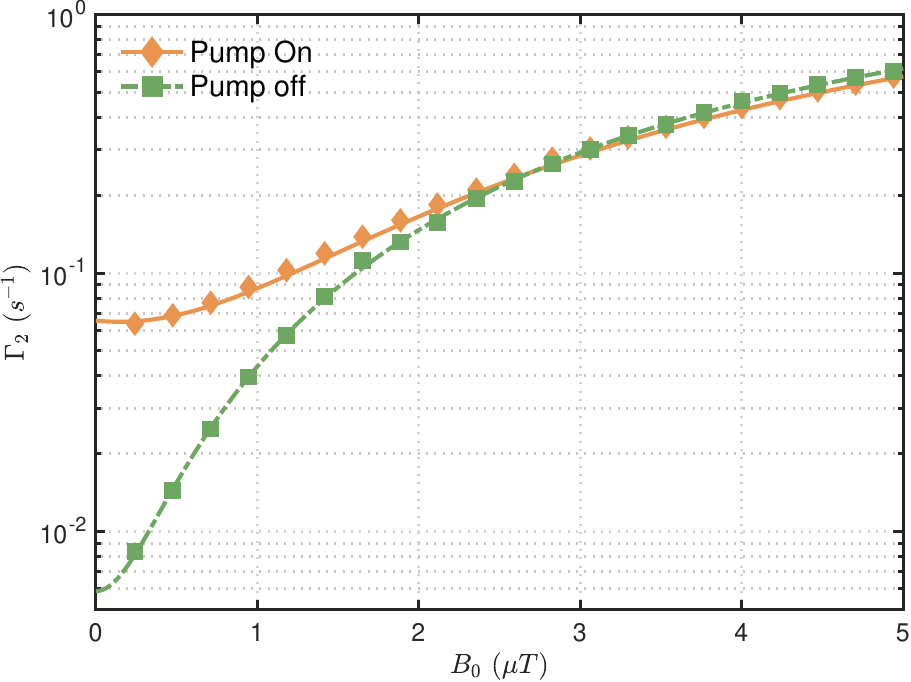}
    % \end{subfigure}
    % \begin{subfigure}{}
    % \includegraphics[width=0.47\linewidth]{CRLB_B0.pdf}
    % \end{subfigure}
    \caption{The transverse relaxation rate versus the strength of the guiding magnetic field with optical pumping on and off, respectively. The yellow diamonds and green squares represent experimental results, while the solid and dashed lines represent theoretical values derived from Eq. (\ref{beta}). 
    % (b) The CRLB of the frequency estimation at different strengths of the guiding magnetic field.
    }
    \label{fig:4}
\end{figure}

The effect of the MECs and light on the precession of the nuclear spin is also related to the strength of the magnetic field.
In Fig. \ref{fig:4}, we demonstrate the transverse relaxation rate $\Gamma_2$ of the ground state $^3$He over the strength of the guiding magnetic field.
The yellow diamonds and green squares represent the relaxation rates extracted from the FID signals, and the solid lines correspond to the theoretical predictions given by the real part of Eq. (\ref{beta}).
When the pump light is blocked, the light-induced fictitious field is zero, i.e., $\delta B_{1/2}=\delta B_{3/2}=0$, then the transverse relaxation rate is reduced to
\begin{equation}
\label{gam2}
    \Gamma_2=\frac{1}{T_r}+\frac{1}{T}(1-\frac{4}{9}\frac{1}{1+(\gamma_m B_0\tau)^2}-\frac{5}{9}\frac{1}{1+(4\gamma_m B_0\tau)^2}).
\end{equation}
Fitting the data marked with green squares based on Eq. (\ref{gam2}), as shown by the dashed curve in Fig. \ref{fig:4}, yields a relaxation time of $T_r = 170$ s, and average MECs process times of $T = 0.37$ s and $\tau = 2.2 \times 10^{-7}$ s for the ground and metastable states, respectively.
The value of $\tau$ obtained here is consistent with that reported in Ref. \cite{dupont1971PRL}. 
The yellow solid curve in Fig. \ref{fig:4} is obtained by substituting the unknown parameters in Eq. (\ref{beta}) with experimentally measured or theoretical values, as listed in Table \ref{tab:1}. 
The experimental results show agreement with the theoretical prediction.
\begin{table}[t!]
    \centering
    \caption{Parameters used for plotting the theoretical curves in Fig. \ref{fig:4}. The radiation decay rate $\gamma_e$ and Doppler linewidth are taken from Ref. \cite{batz2011phd}.}
    \begin{tabular}{cccccc}
        \toprule
        $T_r$ (s) & $T$ (s) & $\tau$ (s) & $I$ (mW/cm$^2$) & $\gamma_e$ (MHz) & 
        $D$ (GHz) \\
        \midrule
        170 & 0.37 & $2.2\times 10^{-7}$ & 100 & $ 10.22$ & $ 1.18$\\
        \bottomrule
    \end{tabular}
    \label{tab:1}
\end{table}
The solid curve shows that the relaxation time approaches a constant value of $1/T_r + 1/T$ at high magnetic fields. 
In the low-field regime where $B_0\ll1/(\gamma_m\tau)\approx 25\ \mathrm{\mu T}$, the MECs-induced relaxation is suppressed, and the relaxation time approaches $T_r$ as the magnetic field tends to zero.
This behavior is similar to the spin-exchange-free region observed in alkali metal magnetometers \cite{allred2002}, although it originates from different mechanisms. 
From the perspective of quantum measurement theory, both relaxation suppression mechanisms can be reinterpreted through the quantum Zeno effect, in which the MECs and spin-exchange collisions can be viewed as measurements on the spin that freezes its evolution in the magnetic field if the measurement frequency exceeds the Larmor frequency \cite{Kominis2008PLA}.
When the pump light is introduced, as shown by the solid curve, it has a significant effect on nuclear relaxation, especially at low magnetic fields.
The relaxation time is significantly shortened to approximately 17 s near zero field, while the effect of light gradually diminishes as the field increases.
This underscores the need to consider the effect of pump light when considering MEOP-polarized $^3$He as a magnetometer, especially for weak field measurements.
% The accuracy of the frequency extracted from the FID signal is crucial in magnetic field measurement. Statistically, the lower limit on the variance of the extracted frequency is dictated by the
% Cramer-Row lower bound (CRLB) \cite{Gemmel2010EPJD}. Figure \ref{fig:4}(b) exhibits the CRLB at different magnetic field strengths. 
% It shows that the extended transverse relaxation time in the low-field regime improves the accuracy of frequency determination. When the pump light is off, the sensitivity achieves nearly a two-order-of-magnitude improvement over the CRLB under low-field conditions, compared to high-field conditions. In terms of magnetic field measurement, a sensitivity level below the picotesla range is achieved in the low-field region. However, the presence of light reduces the sensitivity by nearly one order of magnitude.
\begin{figure}[h!]
    \centering
    \includegraphics[width=1.0\linewidth]{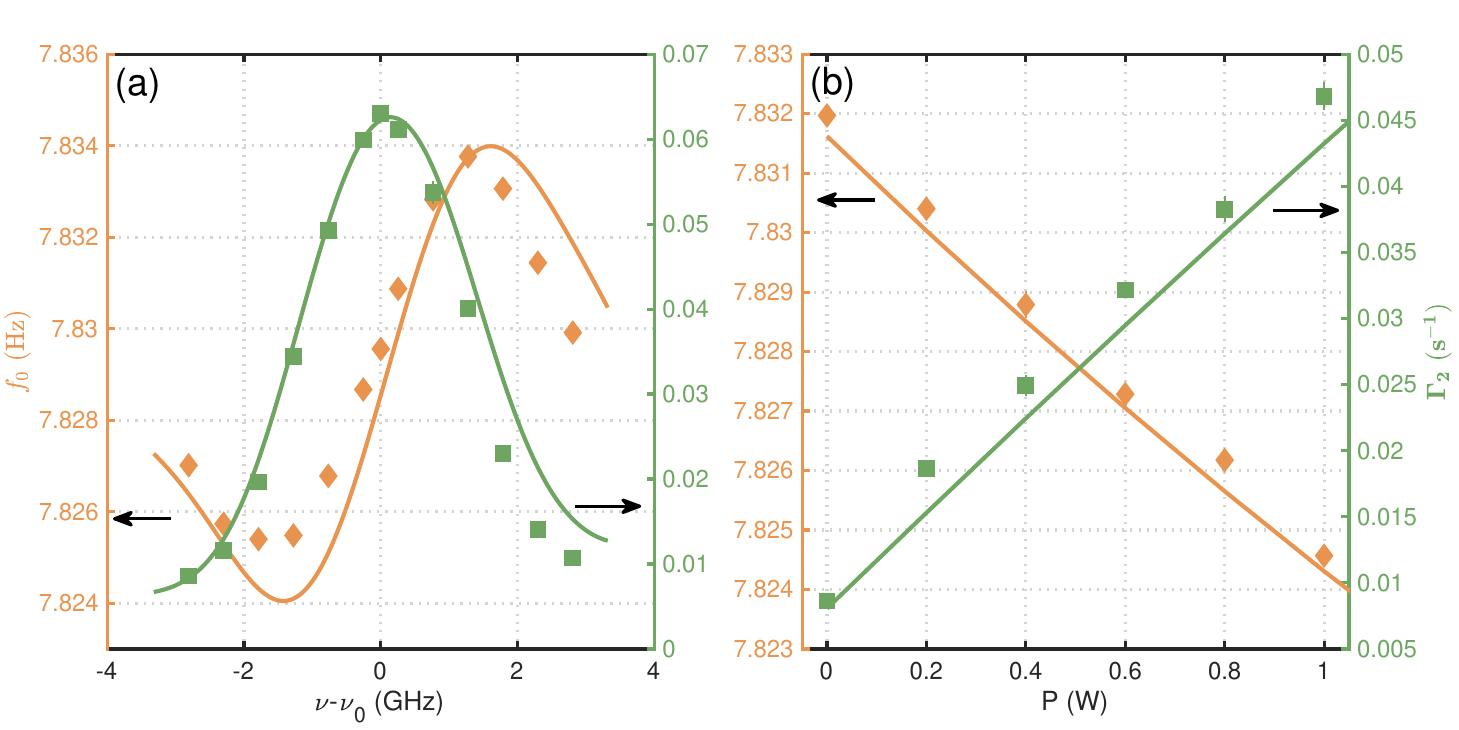}
    \caption{The frequency shift and relaxation induced by the pump light depend on both its wavelength and intensity. Experimental results are shown as green squares and yellow diamonds, while the corresponding theoretical predictions from Eq. (\ref{beta}) are represented by the solid lines. In the left panel, the laser detuning is defined relative to the $C_8$ transition line, with the laser power set to approximately 1 W. In the right panel, the laser wavelength is red-detuned by approximately 1 GHz from the $C_8$ transition.}
    \label{fig:5}
\end{figure}

Finally, we present the dependence of the light-induced frequency shift and relaxation on the detuning and intensity of the pump light, as illustrated in Fig. \ref{fig:5}. 
The yellow diamonds and green squares represent experimental results, while the solid lines correspond to the theoretical predictions given by
Eq. (\ref{beta}).
The theoretical predictions and experimental measurements follow the same trend.
The discrepancy between them may be attributed to uncertainties in the parameters used for the theoretical predictions, such as the laser power, as well as the assumption of a uniform laser power profile across the cell. 
When the laser wavelength is resonant with the transition line (the $C_8$ line), laser-induced relaxation is maximized and diminishes with detuning. 
This is because resonant excitation at the $C_8$ transition leads to maximal broadening of the magnetic resonance linewidth of metastable-state atoms \cite{appelt1999PRA}. 
As a result, nuclear spins in the ground state undergo rapid relaxation after being excited to the metastable state via the MECs.
The relaxation time nearly returns to its value in the absence of the pump light when the laser wavelength is detuned by more than 4 GHz.
The frequency shift as a function of detuning follows a dispersion-like curve. 
As the laser detuning is varied, a relative Larmor frequency shift of nearly 10 mHz is observed.
Note that, compared to the Larmor frequency of 7.832 Hz in the absence of light shift, as indicated in Fig. \ref{fig:1}, the nonzero frequency shift observed at zero detuning results from the laser also exciting the $C_9$ transition.
Finally, since the discharge consistently induces a positive frequency shift \cite{dupont1973b,dupont1973JP}, while the laser-induced shift depends on both the detuning and intensity of the laser, the light shift may be exploited to cancel the total frequency shift.
This enables mitigation of the shift-induced systematic error in the absolute $^3$He magnetometer based on MEOP \cite{Dehmelt1964RSI,lu2025prr}.

\section{Conclusions}
\label{sec:4}
The MECs process has been found to affect the resonance frequency and relaxation of $^3$He nuclear spin precession in the ground state \cite{dupont1973JP,dupont1973b}.
In this work, we show that optical pumping induces an additional frequency shift and enhances the relaxation rate of the nuclear spin via the MECs process.
We develop a theoretical framework to describe the effect of pump light on the ground state of $^3$He.
The pump laser affects the magnetic resonance frequency and linewidth of the metastable state.
Through the MECs process, the ground state is strongly coupled to the metastable state, resulting in an additional frequency shift and enhanced relaxation in the ground state of $^3$He.
In the experiment, we investigate the frequency shift and enhanced relaxation induced by optical pumping by comparing the frequency and relaxation time extracted from the FID signal with and without the pump light.
We studied the light-induced relaxation at different magnetic fields.
The effect of the light shift is appreciable at low magnetic fields but becomes negligible at high magnetic fields.
This highlights the need to carefully consider the pump light when using MEOP-polarized $^3$He as a magnetometer for weak field measurements.
Furthermore, the frequency shift and relaxation as a function of laser detuning and intensity are also examined.
Since the frequency shift caused by the laser depends on its parameters, the light shift may be used to cancel the MECs-induced shift, enabling a precise magnetometer.

Additionally, our work also indicates a bidirectional coupling between the light and the noble gas spins with the aid of MECs. 
On one hand, light shifts the nuclear spin precession frequency; on the other, the polarized nuclear spin modifies the optical susceptibility.
Recently, Katz \textit{et al.} observed that, with the aid of spin-exchange collisions (SEC), the nuclear spin can be effectively excited by light \cite{katz2021SA}. 
Their experiment reveals a coherent bidirectional coupling between light and noble-gas spins that is mediated by alkali atoms.
In view of the similarity of both the MECs and SEC in coupling the electron and nuclear spins \cite{Katz2024arXiv}, we expect that such a kind of coherent coupling is also possible in the MEOP.
Since the MECs process is much faster than the SEC process, it may offer a more efficient means of coupling light to the nuclear spin, facilitating various quantum information applications involving nuclear spins \cite{Dantan2005PRL, Katz2020PRL, Katz2022PRA, Katz2024arXiv}.

\section*{Acknowledgements.} 
We acknowledge support from the National Natural
Science Foundation of China under grant U2230207.

\appendix
\section{The Derivation of $\delta\mathcal{H}$}
\label{app1}
Given that the effective Hamiltonian (\ref{eq1}) is a dot product of two rank-2 tensors, it is convenient to decompose it into the product of irreducible parts using the formula:
\begin{equation}
    (A\cdot B)(C\cdot D)=\sum_{K=0,1,2}(-1)^K \{A\otimes C\}^K\cdot\{B\otimes D\}^K,
\end{equation}
where $A$, $B$, $C$, $D$ are commuting vectors.
In the eigenspace of the hyperfine states $|F_g,m_F\rangle$, the step-by-step derivation of the effective Hamiltonian (\ref{eq1}) proceeds as follows.
{\small
\begin{widetext}
\begin{equation}
\begin{aligned}
\label{deriv}
    \delta\mathcal{H}&=-\frac{I}{2c\varepsilon_0\hbar}Z(\omega;\omega_{F_g})\sum_{K=0,1,2}\sum_{q=-K,\cdots,K} (-1)^{q+K} \{\boldsymbol{\epsilon}^*\otimes \boldsymbol{\epsilon}\}^K_q\alpha^{K}_{-q}\\
    &=-\frac{I}{2c\varepsilon_0\hbar}Z(\omega;\omega_{F_g})\sum_{K=0,1,2} \sum_{q=-K,\cdots,K}\sum_{m_F,m_F'} (-1)^{q+K} \{\boldsymbol{\epsilon}^*\otimes \boldsymbol{\epsilon}\}^K_q\langle F_g,m_F|\alpha^{K}_{-q}| F_g,m'_F\rangle|F_g,m_F\rangle\langle F_g,m'_F|\\
    &=-\frac{I}{2c\varepsilon_0\hbar}Z(\omega;\omega_{F_g})\sum_{K=0,1,2} \sum_{q=-K,\cdots,K}\sum_{m_F,m_F'} (-1)^{q+K} \{\boldsymbol{\epsilon}^*\otimes \boldsymbol{\epsilon}\}^K_q\langle F_g||\alpha^{K}|| F_g\rangle\langle F_g,m_F|F_g,m'_F;K,-q\rangle|F_g,m_F\rangle\langle F_g,m'_F|\\
     &=-\frac{Ie^2}{2c\varepsilon_0\hbar}Z(\omega;\omega_{F_g})\sum_{K=0,1,2} \sum_{q=-K,\cdots,K} \sum_{m_F,m_F'}(-1)^{q+K} \{\boldsymbol{\epsilon}^*\otimes \boldsymbol{\epsilon}\}^K_q\\&(-1)^{K+2F_g}\sqrt{(2F_e+1)(2K+1)}
     \left\{
     \begin{array}{ccc}
         1 & 1 & K \\
         F_g & F_g & F_e
     \end{array}
     \right\}
     \langle F_g||\boldsymbol{r}||F_e\rangle\langle F_e||\boldsymbol{r}||F_g\rangle\langle F_g,m_F|F_g,m'_F;K,-q\rangle|F_g,m_F\rangle\langle F_g,m'_F|\\
     &=-\frac{Ie^2}{2c\varepsilon_0\hbar}Z(\omega;\omega_{F_g})\sum_{K=0,1,2} \sum_{q=-K,\cdots,K} \sum_{m_F,m_F'}(-1)^{q+K} \{\boldsymbol{\epsilon}^*\otimes \boldsymbol{\epsilon}\}^K_q(-1)^{K+2F_g}\sqrt{(2F_e+1)(2K+1)}
     \left\{
     \begin{array}{ccc}
         1 & 1 & K \\
         F_g & F_g & F_e
     \end{array}
     \right\}\\&(-1)^{F_e-F_g}\sqrt{\frac{2F_g+1}{2F_e+1}}
     |\langle F_g||\boldsymbol{r}||F_e\rangle|^2\langle F_g,m_F|F_g,m'_F;K,-q\rangle|F_g,m_F\rangle\langle F_g,m'_F| \\
     &=-\frac{Ie^2}{2c\varepsilon_0\hbar}Z(\omega;\omega_{F_g})\sum_{K=0,1,2} \sum_{q=-K,\cdots,K} \sum_{m_F,m_F'}(-1)^{q+K} \{\boldsymbol{\epsilon}^*\otimes \boldsymbol{\epsilon}\}^K_q\\
     &(-1)^{K+2F_g}\sqrt{(2F_e+1)(2K+1)}
     \left\{
     \begin{array}{ccc}
         1 & 1 & K \\
         F_g & F_g & F_e
     \end{array}
     \right\}(-1)^{F_e-F_g}\sqrt{\frac{2F_g+1}{2F_e+1}}\\
     &\Big((-1)^{F_e+J_g+1+I}\sqrt{(2F_e+1)(2J_g+1)}
     \left\{
     \begin{array}{ccc}
         J_g & J_e & 1 \\
         F_e & F_g & I
     \end{array}
     \right\}
     \Big)^2|\langle J_g||\boldsymbol{r}||J_e\rangle|^2\langle F_g,m_F|F_g,m'_F;K,-q\rangle|F_g,m_F\rangle\langle F_g,m'_F|\\
     &=-\frac{Ie^2f_{ge}}{4m_e\omega_{F_g}c\varepsilon_0}Z(\omega;\omega_{F_g})\sum_{K=0,1,2} \sum_{q=-K,\cdots,K} \sum_{m_F,m_F'}(-1)^q \{\boldsymbol{\epsilon}^*\otimes \boldsymbol{\epsilon}\}^K_q\\&\sqrt{(2F_e+1)(2K+1)}
     \left\{
     \begin{array}{ccc}
         1 & 1 & K \\
         F_g & F_g & F_e
     \end{array}
     \right\}(-1)^{F_e+F_g}\sqrt{\frac{2F_g+1}{2F_e+1}}\\
     &(2J_e+1)(2J_g+1)\Big((-1)^{F_e+J_g+1+I}\sqrt{(2F_e+1)(2J_g+1)}
     \left\{
     \begin{array}{ccc}
         J_g & J_e & 1 \\
         F_e & F_g & I
     \end{array}
     \right\}
     \Big)^2\langle F_g,m_F|F_g,m'_F;K,-q\rangle|F_g,m_F\rangle\langle F_g,m'_F|,
\end{aligned}
\end{equation}
\end{widetext}
}
\noindent where $f_{eg}=\frac{2m_e\omega_{F_g}}{\hbar(2J_e+1)(2J_g+1)}|\langle J_g||\boldsymbol{r}||J_e\rangle|^2$ is the oscillator strength.
Note that the coherence between the two hyperfine levels in the metastable is neglected due to the MECs process \cite{courtade2002EPJD}.
Necessary formulas used in the derivation can be found in Ref. \cite{steck2007quantum}. By substituting the specific values of the angular momenta and evaluating the 6-$j$ symbols using Mathematica, we obtain the reduced Hamiltonian.

The scalar (K=0) part is:
\begin{equation}
    \delta\mathcal{H}^{(0)}=-\frac{Ie^2f_{ge}}{4m_ec\varepsilon_0}(\frac{Z(\omega;\omega_8)}{\omega_8}P_{F=1/2}+\frac{Z(\omega;\omega_9)}{\omega_9}P_{F=3/2}),
\end{equation}
where $P_{F_i}=\sum_{m}|F_i,m_{F_i}\rangle\langle F_i,m_{F_i}|$ is the projector operator onto the hyperfine structure.

The vector (K=1) part is:
\begin{equation}
\begin{aligned}
    \delta\mathcal{H}^{(1)}
    %&=\frac{3If_{ge}\hbar}{4m_e\omega_0c\varepsilon_0}(\frac{2\sqrt{2}}{3}\frac{i}{\sqrt{2}}(\boldsymbol{\epsilon}^*\times\boldsymbol{\epsilon})\cdot \boldsymbol{F}_{1/2}+\frac{\sqrt{2}}{3}\frac{i}{\sqrt{2}}(\boldsymbol{\epsilon}^*\times\boldsymbol{\epsilon})\cdot \boldsymbol{F}_{3/2})\\
    &=\frac{Ie^2f_{ge}}{4m_ec\varepsilon_0}(2\frac{Z(\omega;\omega_8)}{\omega_8}\boldsymbol{s}\cdot \boldsymbol{F}_{1/2}+\frac{Z(\omega;\omega_9)}{\omega_9}\boldsymbol{s}\cdot \boldsymbol{F}_{3/2}),
    \end{aligned}
\end{equation}
where $\boldsymbol{s}=\frac{(\boldsymbol{\epsilon}^*\times\boldsymbol{\epsilon})}{i}$ is the mean spin of the photon.

The tensor (K=2) part is:
\begin{equation}
\begin{aligned}
    \delta\mathcal{H}^{(2)}&=\frac{Ie^2f_{ge}}{12m_ec\varepsilon_0}\frac{Z(\omega;\omega_9)}{\omega_9}[3(\boldsymbol{\epsilon}\cdot \boldsymbol{F}_{3/2})(\boldsymbol{\epsilon}^*\cdot \boldsymbol{F}_{3/2})\\
    &-\boldsymbol{F}_{3/2}^2].
\end{aligned}
\end{equation}

\end{document}